\documentclass[twocolumn,showpacs,preprintnumbers,amssymb]{revtex4}

\usepackage{graphicx}
\usepackage{epsfig}
\usepackage{dcolumn}
\usepackage{bm}
\def\gsim{\lower0.5ex\hbox{$\:\buildrel >\over\sim\:$}}
\def\lsim{\lower0.5ex\hbox{$\:\buildrel <\over\sim\:$}}

\def \n{\noindent}

\let\d=\delta
\let\e=\epsilon\let\g=\gamma

\let\m=\mu\let\n=\nu
\let\s=\sigma

\newcommand{\be}{\begin{equation}}
\newcommand{\ee}{\end{equation}}
\newcommand{\bea}{\begin{eqnarray}}
\newcommand{\eea}{\end{eqnarray}}

\newcommand{\nbox}{{\,\lower0.9pt\vbox{\hrule \hbox{\vrule height 0.2 cm
\hskip 0.2 cm \vrule height 0.2 cm}\hrule}\,}}

\begin{document}

\preprint{UCI-TR-2008-25}

\title{
On the Decay of Unparticles}

\author{Arvind Rajaraman
}
\email{arajaram@uci.edu}
\affiliation{
Department of Physics and Astronomy, University of California,
Irvine, CA 92697, USA}

\date{\today}

\begin{abstract}
We show that when the unparticle sector is coupled to the Standard
Model, unparticle excitations can decay to Standard Model
particles. This radically modifies the signals of unparticle
production. We present a method for the calculation of
the decay lifetimes of unparticles.
In a particular model, we show that depending
on their lifetime,  unparticles can manifest themselves through
monojets, delayed events or prompt decays.
\end{abstract}


\maketitle

\section{Introduction} Recently Georgi
\cite{Georgi:2007ek,{Georgi:2007si}} has suggested the possibility
that there is a new, almost conformal, sector coupled to the
Standard Model through interactions of the form $O_{\cal
U}O_{SM}$, where $O_{\cal U}$ is an operator of the conformal
field theory, and $O_{SM}$ is a Standard Model operator. This
conformal sector can have experimental signals radically different
from those of normal particles, and hence was dubbed the
unparticle sector. 
There has since been a great deal of work studying both the
theoretical and experimental aspects of the unparticle sector.

Unparticles manifest themselves through two kinds of processes.
In the first class of processes,
the unparticle propagator 
mediates
processes like $q\bar{q}\rightarrow l^+l^-$ or $gg\rightarrow
l^+l^-$. This can lead to striking signals at colliders
\cite{Cheung:2007ap,{Bander:2007nd},{Mureika:2007nc},{Rizzo:2007xr},{Mathews:2007hr},{Cheung:2007zza},{Feng:2008ae}}.
 The propagator can also affect precision
measurements~\cite{Luo:2007bq,{Liao:2007bx},{Lenz:2007nj}}.

Another class of processes involves the real production of
unparticles (e.g. \cite{{Georgi:2007ek},Cheung:2007zza,
{Chen:2007zy}}). The peculiar phase space properties of
unparticles then leads to
distinctive signals. In this 
class of analyses, it is
 usually assumed that the unparticle escapes undetected,
 and is manifested as missing energy. Examples of such
 processes are $e^+ e^-\rightarrow \gamma O_{\cal U}$
 and   $gg\rightarrow g O_{\cal U}$ which have been assumed
 to lead to monophoton production
 and monojet production respectively~\cite{{Cheung:2007zza}}.

Here we revisit this latter class of processes.
We show that
in
fact the produced unparticles can decay back to Standard Model
particles, just like  normal resonances (similar comments have been made in
\cite{Strassler:2008bv}). {\it Therefore,
in general, unparticles are not characterized by missing energy signals}.
This fact
will drastically modify both current
constraints and future discovery prospects of unparticles.

Our main result in this note 
is a method for calculating the decay rates
of  unparticles, at least for scalar unparticles in the range $1\le d\le 1.5$ (the
restriction is explained below.)
 We also apply this formalism to a
specific model, where the unparticle is coupled to massless vector
bosons.
We
find a large class of possible signals, depending on the unparticle
lifetime.
If the lifetime is very long, the unparticles
will escape the detector, and  we will
return to the situation where unparticles are characterized by missing energy.
If the lifetime is  short, we get prompt decays.
Most interestingly, for a broad
class of parameters, the unparticles may travel a macroscopic
distance before decaying, and this will lead to delayed signals
and displaced vertices.

We close with a discussion of other situations to which our
formalism can be applied.
We emphasize that our results will affect many analyses in the
unparticle literature
involving real unparticles, such as heavy
 quark decay to unparticles. We will leave these and
 other questions to future work.

\section{Unparticle decays}

The propagator for unparticles in an exactly conformal theory
is fixed by conformal invariance to be
~\cite{Georgi:2007si,Cheung:2007zza}
\begin{equation}
iB_dD_0(p^2)=   i B_d (p^2)^{d-2}
\end{equation}
where
\begin{equation}
    B_d \equiv A_d\, \frac{\left(e^{-i \pi} \right)^{d-2}}
    {2 \sin d \pi}\, , \hspace{0.3cm}
   A_d \equiv \frac{16\, \pi^{5/2}\, \Gamma (d+ \frac{1}{2})}
    {(2 \pi)^{2d}\, \Gamma(d-1)\, \Gamma (2d)}\, ,
\end{equation}
and $1\le d< 2$ is the dimension of the operator $O_{\cal U}$.

The coupling to the Standard Model modifies the propagator.
The most important coupling is the relevant operator $O_{\cal U} H^2$,
which breaks conformal
invariance~\cite{Fox:2007sy,{Delgado:2007dx},{Kikuchi:2007qd}} and introduces a
scale $\mu$ into the conformal field theory. The precise
modification to the propagator is model dependent~\cite{Delgado:2008rq}; here we will
use a simple model proposed in ~\cite{Fox:2007sy}, where the
modified propagator is
\begin{equation}
iB_dD(p^2)=   i B_d (p^2-\mu^2)^{d-2}
\end{equation}
which has the feature that at high $p^2$ we
recover the original propagator, and for $p^2<\mu^2$,
the propagator vanishes since there are no states.

It appears that the propagator is complex. This can be understood by
writing the propagator using a dispersion
relation as
\bea
iB_dD(p^2)=
i {A_d\over 2\pi}\int_0^\infty dM^2 {\rho(M^2)\over p^2-\mu^2-M^2+i\e}
\eea
with $\rho(M^2)=(M^2)^{d-2}$.
In this representation it is manifest that the complex nature of the
propagator is not related to a decay, but  rather is because
 the propagator is a sum
over resonances \cite{Stephanov:2007ry}.

There are also other couplings to other fields through irrelevant operators like
$O_{\cal U} (F_{\mu\nu})^2,O_{\cal U}\bar{\psi}\psi $
(along with $O_{\cal U}H^2$). These will  modify the propagator
through loop contributions.
We can resum the contributions 
to obtain the full propagator
\bea
\int e^{ipx}\langle 0| T\left( O_{\cal U}(x)O_{\cal U}(0)\right)|0\rangle d^4x~~~~~~~
~~~~~~~~~~~~~~~\nonumber
\\
=iB_dD(p^2)+iB_d^2D(p^2)\Sigma(p^2)D(p^2)+...
\\
\equiv {iB_d\over
(p^2-\mu^2)^{2-d}-B_d\Sigma(p^2)}\nonumber
\eea
where the loop diagram is $-i\Sigma(p^2)$.

Our main concern will be with the effects on the unparticle of $\Sigma(p^2)$.
In a particle propagator (i.e. $d=1$), an imaginary part of
$\Sigma(p^2)$ leads to a width for the particle, signalling that it
can decay. The decay rate is related to the imaginary part of the
propagator through the Cutkosky rules. We expect a similar effect for
unparticles.

To make this explicit, we will  attempt to express the modified propagator
as a dispersion integral.
We will first do this assuming that $(p^2-\mu^2)^{2-d}\gg |B_d|\Sigma(p^2)$, so
that the
loop term 
can be treated as a perturbation. In this regime, the
propagator is approximately
\bea
{iB_d\over
(p^2-\mu^2-B_d\Sigma(p^2){(p^2-\mu^2)^{d-1}\over
2-d})^{2-d}}~~~~~~~~~~~~~~~~~~~~~~~~\nonumber
\\
=i {A_d\over 2\pi}\int_0^\infty dM^2 {(M^2)^{d-2}
\over p^2-\mu^2-M^2-B_d\Sigma(p^2){(p^2-\mu^2)^{d-1}\over 2-d}}
\eea
If $B_d\Sigma(p^2)$ develops an imaginary piece, this will appear as a width
for the resonances,
leading to a decay, in analogy with particles. Note that $\rho(M^2)$
is necessarily real, so the inclusion of
the width is mandatory.

There is unfortunately a problem with this interpretation.
If $\Sigma(p^2)$ is real and nonzero,  $B_d\Sigma(p^2)$ can be complex.
This would lead to a width for the unparticle even if the resonance
is forbidden to
decay to Standard Model particles.

Our resolution to this puzzle is that  $\Sigma(p^2)$ is not
allowed to have a  nonzero real part; i.e. we need to add
counterterms so that the real part is canceled (the necessity of
such counterterms in other situations was pointed out in
\cite{Grinstein:2008qk}). In a normal particle theory, this would
be impossible, since the counterterms are local, but in a nonlocal
unparticle theory, there is no restriction coming from locality.
The counterterms are now constrained by unitarity rather than
locality. If this resolution is correct, then we  only need to
consider the case when $\Sigma(p^2)=-i\Sigma_I(p^2)$ is pure
imaginary. This occurs when the unparticle is kinematically
allowed to decay to standard Model particles, and will produce a
width for the unparticle, as required by unitarity.

The width of the resonance of mass $M$ may then be
read off 
to
be
\bea
\Gamma(M)={\Sigma_I(M^2)\over (2-d)M}(M^2-\mu^2)^{d-1}{A_d\over 2}{\cot(\pi d)}
\eea

We thus return to the 
description \cite{Stephanov:2007ry} of
the unparticle as a set of resonances
with a continuous distribution of masses; the new feature is that these
resonances decay, with a lifetime
$\Gamma^{-1}(M)$. The position space propagator
then has the form
\bea
\langle 0| T( O_{\cal U}(x)O_{\cal U}(0))|0\rangle\propto
\int dM^2 \rho(M^2) e^{-M\Gamma(M)t}
\eea
which explicitly shows that the unparticle propagator has a decay.

We note that our expression does not work if $d>1.5$, where the
width is  negative.
 This is
 a failure of our perturbation expansion. We have not been
able to find a way to extend the deconstructed expression to the region
$d>1.5$; henceforth we will only consider the situation where
$d\le 1.5$.

This perturbation expansion also fails if $\Sigma_I(p^2)$ is
too large. 
Since the loop contribution
is small, this only happens very close to the
mass gap
at the point $p^2\leq p_0^2$ with
$(p_0^2-\mu^2)^{2-d}= |B_d|\Sigma_I(\mu^2)$.
In this regime, we may  approximate
\bea
{iB_d\over (p^2-\mu^2)^{2-d}-iB_d\Sigma_I(p^2)}~~~~~~~~~~~~~~~~
~~~~~~~~~\nonumber
\\
\simeq {i|B_d|\over {p^2-\mu^2\over (p_0^2-\mu^2)^{d-1}}
-i|B_d|\Sigma_I(\mu^2)}
\eea
which is of the deconstructed form with
\bea
\rho(M^2)={2\pi\over A_d}{|B_d|
}(p_0^2-\mu^2)^{d-1}\d(M^2-\mu^2)~~~
\\
\Gamma(M)
=\Sigma_I(\mu^2){(p_0^2-\mu^2)^{d-1}\over {\mu}}
{|B_d|
}~~~~~~~~~~~~
\eea
It would be interesting to find a more
accurate representation of the propagator for this regime.

To summarize, we have shown that unparticles have decays, just like
normal particles. The unparticle can be regarded
a sum over several particle propagators,
where the particles have a continuously distributed mass $M$,
and a width $\Gamma(M)$. The width is related to the imaginary part
of the loop correction as required by unitarity.

Before turning to the experimental consequences, we briefly comment on
previous arguments in the literature  that
unparticles do not decay. The argument comes from looking at the
deconstructed form of the propagator; each resonance of mass $M$ has an
infinitesimal coupling which would appear to preclude decay.
The resolution to this paradox is that the mass of
the unparticle is not a well defined quantity.
 The unparticle should be treated as having a fixed momentum, and one should
 sum over all resonance masses, keeping $p^2$ fixed.
 The sum over the infinite set of resonances compensates for the
 infinitesimal coupling, and we get a finite decay
 rate.

 \section{Experimental signatures}

We will now examine how this decay affects the experimental signals of
unparticles. We shall here consider in detail one
particular model, where the unparticle couples
mainly to massless vector bosons,
and analyze the signatures of such unparticles at the LHC.
We take the couplings to be
\bea
L_{int}={O_{\cal U}F_{\m\n}F^{\m\n}\over \Lambda_F^d}+{O_{\cal U}
G_{\m\n}G^{\m\n}\over \Lambda_G^d}
\eea
where $F_{\m\n}, G_{\m\n}$ are the electromagnetic and color field
strengths respectively, and
$\Lambda_F, \Lambda_G$ are scales parametrizing the
couplings. We will take for simplicity
 $\Lambda_F\sim \Lambda_G=\Lambda$.

We first calculate the widths using the formalism above.
There is a crossover point for these widths,
which occurs at a scale $p_0$ set by
\bea
\left({p_0^2-\mu^2\over \Lambda^2}\right)^{2-d}= |B_d|{\m^4\over 2\pi\Lambda^{4}}
\eea
For $M^2 > p_0^2$ we find
\bea \rho(M^2)=
(M^2-\mu^2)^{d-2}
~~~~~~~~~~~~~~~~~~~~~~~~~
\\
\Gamma(M)
=\left({M^2-\m^2\over
\Lambda^2}\right)^{d-1}{M^3\over 4\pi\Lambda^2}{A_d\cot(\pi d)\over (2-d)}\eea
while
for $M^2<p_0^2$ we have
\bea
\rho(M^2) = {2\pi\over A_d}{|B_d|
}(p_0^2-\mu^2)^{d-1}\d(M^2-\mu^2)~~
\\
\Gamma(M)
= \left({p_0^2-\mu^2\over
\Lambda^2}\right)^{d-1}{\m^3\over 2\pi\Lambda^2}{|B_d|
}~~~~~~~~~~ \eea

Unparticles with these interactions can be produced at
colliders 
through gluon fusion,
in processes like $gg\rightarrow O_U, gg\rightarrow gO_U$.
If one does not consider unparticle decay, the unparticle
will escape, and this leads to missing energy signals;
in particular the process $gg\rightarrow gO_U$ leads to monojet signals.

However, if unparticle decay is considered, then the signals are very different.
The unparticle can decay through the processes $O_{\cal U}\rightarrow gg$ and
 $O_{\cal U}\rightarrow \g\g$, leading to  multijet events, or events with
  two photons plus  jets.
 In particular, there will be few or no
 missing energy events or monojets, unless the lifetime
 is very long.

We henceforth focus on the process $gg\rightarrow gO_{\cal U}$.
The cross-section for
this process is  found to be
\bea
{d\s\over d\hat{t}dM^2}={1\over 16\pi \Lambda^{2d}{\hat{s}}^2}
{A_d\over 2\pi}\rho(M^2)
|{\cal M}^2|
\eea
with
\bea
|{\cal M}^2|={1536\pi \alpha_s\over 4.8.8}{(M^2)^4+(\hat{s}^2+\hat{t}^2
+\hat{u}^2)^2\over 
\hat{s}\hat{t}\hat{u}}
\eea

The produced unparticles will decay either to gluons or
photons.
The resulting signals are of several types.

{\it a. Monojets:}  If the unparticle decays outside the detector, it is
effectively missing energy,
and we get a monojet signal.

To estimate the number of monojet events, we need the
 cross-section for events in which the unparticle decays
 after moving a  distance $r$. This is
\bea
\s=
\int dM^2 d\hat{t} {d\s\over dM^2
d\hat{t}}exp(-{M\Gamma(M) r\over p})
\eea
We assume the detector size to be 10 cm.
Therefore, the number of monojets is the number
of events where the  displacement of the vertex is greater than
10 cm.
We will in addition
require the gluon jet to have $E>100$ GeV,
and to have rapidity
$\eta<2.5$.

\begin{figure}[t]
\centering
\includegraphics[width=8cm]{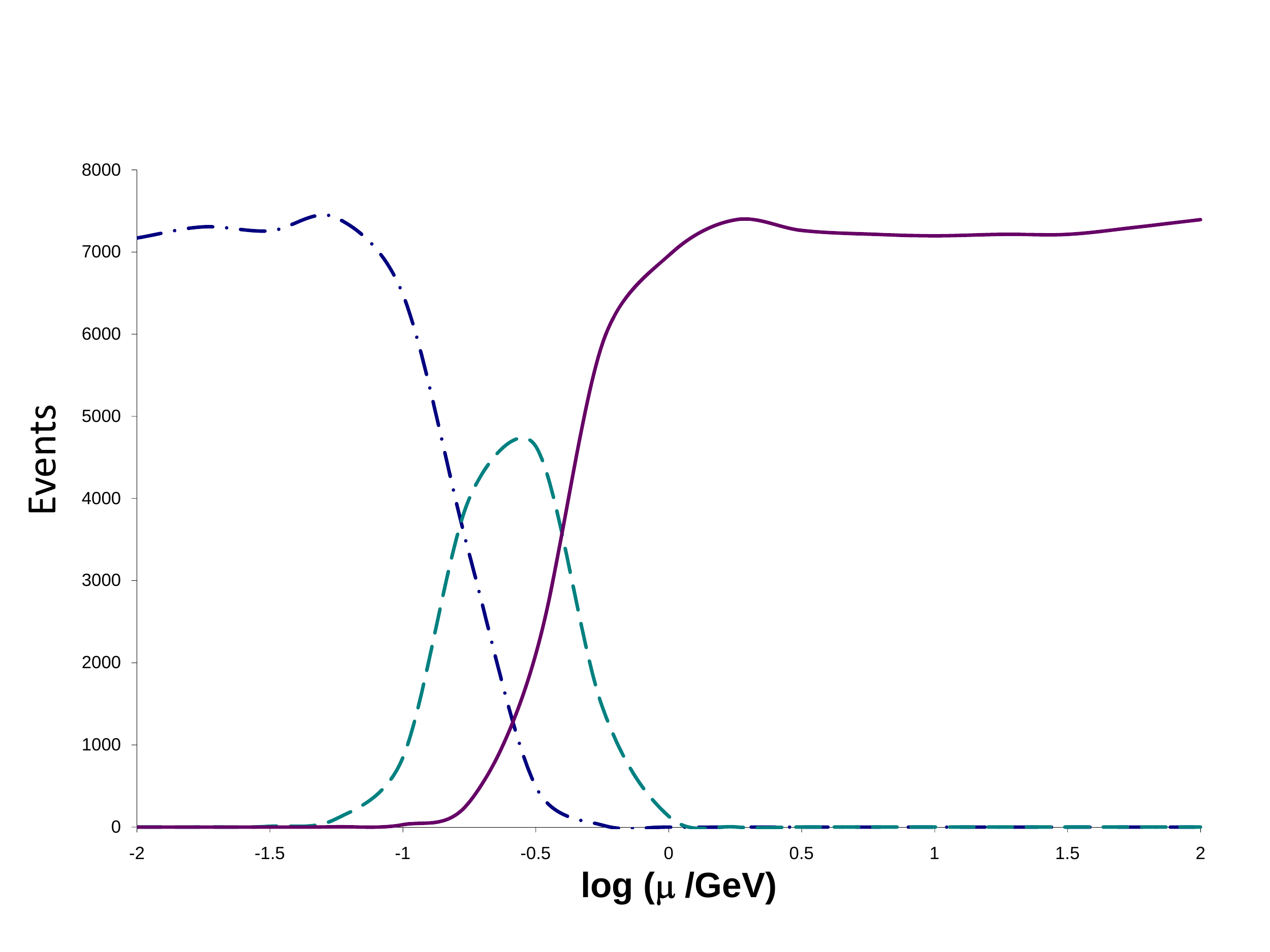}
\caption{Number of events with $10$ fb$^{-1}$ of LHC data as a function of $\mu$.
The solid (red) line corresponds to the number of prompt events,
dot-dashed (blue) corresponds to the number of monojet events, and
dashed (green) is the number of delayed events. We have taken $d=1.1$ and
$\Lambda=10000$ GeV.}
\label{d11fig}
\end{figure}

{\it b. Delayed jets/photons}
: A more striking signal is provided by the  situation where
the unparticle decays before exiting the detector.
The decay will produce either photons or gluons which are detectable,
and with a time delay given by the
lifetime of the unparticle.
The total cross-section for such events where the unparticle has a  lifetime
$t$ is
\bea
\s=
\int dM^2 d\hat{t} {d\s\over dM^2
d\hat{t}}exp(-{M\Gamma(M) t\over E})
\eea
We note that the unparticle will usually be strongly boosted, and the
decay products will then be almost collinear and will
appear as a single photon/jet. The signal will be a delayed
photon/jet, accompanied by a hard jet.

The  detection threshold is set by the
timing resolution of the detectors.
According to \cite{Vigano:2006zz}, the ATLAS calorimeter has a timing resolution of
100 ps, and  the CMS EM calorimeter has a comparable resolution.
We
therefore require the time delay to be greater than 100 ps.
 We will in addition
require the gluon jet to have $E>100$ GeV,
and to have rapidity
$\eta<2.5$.


{\it c. Prompt decays}: If the lifetime of the unparticle is less than 100ps, the
decay is prompt. We then get  two photons with an extra hard jet.
These are similar to the virtual unparticle processes.

\begin{figure}[t]
\centering
\includegraphics[width=8cm]{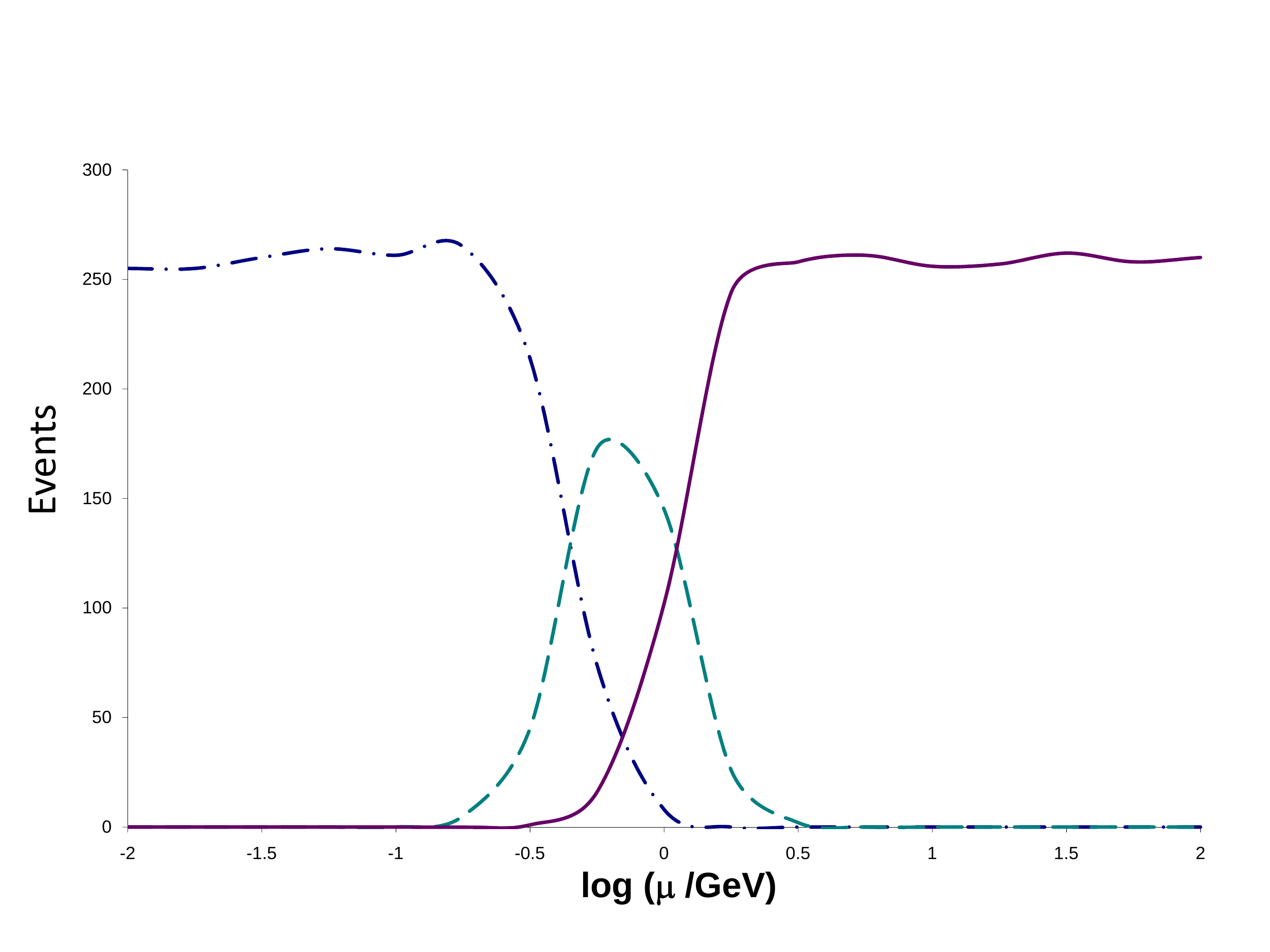}
\caption{Similar to Fig. 1, but for $d=1.4$.}
\label{d14fig}
\end{figure}

We now calculate the number of signal events of
each type as a function of $\mu$. We will assume $10$ fb$^{-1}$ of LHC data.
The numbers of such events is shown in Fig. \ref{d11fig} for $d=1.1$ and
Fig. \ref{d14fig} for $d=1.4$.
We see that for larger values of $\mu$, the decays are
almost all prompt. For small $\mu$, more unparticles with a
long lifetime can be produced, and we get a large number of monojets.
In the intermediate range ($\mu\sim 1$ GeV), we find a significant number of
delayed events. This provides a new type of signal
of unparticles.

We have ignored issues of efficiencies and backgrounds; these
must of course
be included in a realistic analysis.

\section{Summary and Future Directions}

We have shown that unparticles can decay to standard model
particles, and found an expression for their decay rate.
We have applied this to a particular model, and shown
that such effects can have striking signals. In
particular for a range of parameters, we can have delayed
events, where the unparticle travels a significant time
before decaying.

These results drastically affect many other unparticle analyses in the literature.
For example, the seminal paper \cite{Georgi:2007ek}
considered the coupling $O_{\cal U}\bar{t}u$ which
can mediate
top decay through the process $t\rightarrow uO_{\cal U} $. If the unparticle does
not decay, this would be observed as a missing energy+jet signal. However,
 the unparticle can decay through $O_{\cal U}\rightarrow bWu$, and this implies
that a different possible decay mode is  $t\rightarrow ubWu$. Furthermore,
if the unparticle has a significant
lifetime, the unparticle decay products may come from a significantly
displaced vertex.
It would be very
interesting to examine the experimental constraints on this process, and
on similar decay modes like $b\rightarrow  sO_{\cal U}$.

Finally, it would be very interesting to understand how to extend
our formulae to the case $d>1.5$. We will leave this for future work.

{\it Acknowledgments:} We thank T. C. Yuan for pointing out an error in the original manuscript.
This work is supported in part by
NSF Grants No.~PHY--0354993 and PHY--0653656.

\end{document}